\documentclass[twocolumn,prb,superscriptaddress]{revtex4}% Physical Review B
\usepackage{graphicx}%
\usepackage{dcolumn}
\usepackage{amsmath}

\makeatletter
\def\btt#1{\texttt{\@backslashchar#1}}%
\DeclareRobustCommand\bblash{\btt{\@backslashchar}}%
\makeatother

\begin{document}

\preprint{typeset using REV4}

\title{Structural, transport, and thermal properties of single
crystalline type-\setcounter{enumi}{8}\Roman{enumi} clathrate Ba$_8$Ga$_{16}$Sn$_{30}$ }% Force line breaks with \\

\author{D. Huo}
\affiliation{College of Materials Science and Engineering, Donghua
University, Shanghai 200051, China} \affiliation{Department of
Quantum Matter, ADSM, Hiroshima University, Higashi-Hiroshima
739-8530, Japan}
\author{T. Sakata}
\affiliation{Department of Quantum Matter, ADSM, Hiroshima
University, Higashi-Hiroshima 739-8530, Japan}
\author{T. Sasakawa}
\affiliation{Department of Quantum Matter, ADSM, Hiroshima
University, Higashi-Hiroshima 739-8530, Japan}
\author{M. A. Avila}
\affiliation{Department of Quantum Matter, ADSM, Hiroshima
University, Higashi-Hiroshima 739-8530, Japan}
\author{M. Tsubota}
\affiliation{Department of Quantum Matter, ADSM, Hiroshima
University, Higashi-Hiroshima 739-8530, Japan}
\author{F. Iga}
\affiliation{Department of Quantum Matter, ADSM, Hiroshima
University, Higashi-Hiroshima 739-8530, Japan}
\author{H. Fukuoka}
\affiliation{Department of Applied Chemistry, Hiroshima University,
Higashi-Hiroshima 739-8527, Japan}
\author{S. Yamanaka}
\affiliation{Department of Applied Chemistry, Hiroshima University,
Higashi-Hiroshima 739-8527, Japan}
\author{S. Aoyagi}
\affiliation{The Japan Synchtron Radiation Research Institute,
SPring-8, Hyogo 679-5108, Japan}
\author{T. Takabatake}
\affiliation{Department of Quantum Matter, ADSM, Hiroshima
University, Higashi-Hiroshima 739-8530, Japan}

%\date{\today}% It is always \today, today, but you may specify any date with \date.

\begin{abstract}
We report the electrical resistivity $\rho$, Hall coefficient
$R_{\rm H}$, thermoelectric power $S$, specific heat $C$, and
thermal conductivity $\kappa$ on single crystals of the
type-\setcounter{enumi}{8}\Roman{enumi} clathrate
Ba$_8$Ga$_{16}$Sn$_{30}$ grown from Sn-flux. Negative $S$ and
$R_{\rm H}$ over a wide temperature range indicate that electrons
dominate electrical transport properties. Both $\rho$(T) and $S(T)$
show typical behavior of a heavily doped semiconductor. The absolute
value of $S$ increases monotonically to 243 $\mu$V/K with increasing
temperature up to 550 K. The large $S$ may originate from the low
carrier concentration $n$=3.7$\times$10$^{19}$ cm$^{-3}$. Hall
mobility $\mu_{\rm H}$ shows a maximum of 62 cm$^2$/Vs around 70 K.
The analysis of temperature dependence of $\mu_{\rm H}$ suggests a
crossover of dominant scattering mechanism from ionized impurity to
acoustic phonon scattering with increasing temperature. The
existence of local vibration modes of Ba atoms in cages composed of
Ga and Sn atoms is evidenced by analysis of experimental data of
structural refinement and specific heat, which give an Einstein
temperature of 50 K and a Debye temperature of 200 K. This local
vibration of Ba atoms should be responsible for the low thermal
conductivity (1.1 W/m K at 150 K). The potential of
type-\setcounter{enumi}{8}\Roman{enumi} clathrate compounds for
thermoelectric application is discussed.
\end{abstract}

\pacs{65.40.-b, 72.20.Pa, 72.15.-v}% PACS, the Physics and Astronomy Classification Scheme.

 \maketitle

%\tableofcontents

\section{Introduction}
Semiconducting clathrate compounds are attracting considerable
attention because of their potential applications in
thermoelectrics.\cite{1} These compounds consist of face-shared
polyhedral cages (formed by Si, Ge, Sn, and/or Ga) filled with
alkali-metal, alkaline-earth and/or rare-earth atoms. The most
pronounced feature of clathrate compounds is their very low lattice
thermal conductivity $\kappa_{\rm L}$($\sim$ 1W/m K at room
temperature). Some compounds even show glasslike
temperature-dependent thermal conductivity, although they
crystallize in well-defined structures. These classes of compound
are good candidates to fulfill the phonon glass electron crystals
(PGECs) concept,\cite{2} which is a guideline to search for high
performance thermoelectric materials with the compatibility of low
thermal conductivity and high electrical conductivity. The
thermoelectric performance of a material at a given operation
temperature $T$ is characterized by the dimensionless figure of
merit $ZT$, which is defined as $ZT$=$S^2T/\rho(\kappa_{\rm
L}+\kappa_{\rm e}$), where $S$, $\rho$, $\kappa_{\rm e}$ are the
thermoelectric power, electrical resistivity, and electronic thermal
conductivity of the material, respectively. A higher energy
conversion efficiency demands a large $ZT$. However, $ZT$ has been
limited to unity for several decades although much effort has been
made to increase it.

Recently, within the spirit of PGEC concept, open structured
compounds such as filled skutterudites and clathrates have been
extensively investigated due to their low $\kappa_{\rm L}$, which
leads to a much-enhanced $ZT$.\cite{1,3,4} The reduction of thermal
conductivity for these compounds is believed resultant from the
local vibrations (rattling) of the guest atoms encapsulated in
oversized cages. The heat-carrying phonons are scattered effectively
by the rattling of these guest atoms. However, the mechanisms
responsible for some clathrate compounds showing glasslike
$\kappa$(T) at low temperatures remain an open issue. Based on the
experimental results of neutron scattering and ultrasonic
attenuation on single crystals of $X_8$Ga$_{16}$Ge$_{30}$($X$ = Ba,
Sr, Eu),\cite{5,6} it was concluded that the scattering of phonons
from tunneling states is responsible for the glasslike $\kappa$(T)
of (Sr/Eu)$_8$Ga$_{16}$Ge$_{30}$ in addition to the scattering from
the rattling guest atoms. The absence of glasslike $\kappa(T)$ for
the $n$-type Ba$_8$Ga$_{16}$Ge$_{30}$ sample was attributed to a
very low density of tunneling states, if any. On the other hand,
Bentien \emph{et al}.\cite{7} recently reported a glasslike
$\kappa$(T) of a Ga-rich $p$-type Ba$_8$Ga$_{16}$Ge$_{30}$. They
discussed the difference in thermal conductivity between the two
types of samples and pointed out that glasslike $\kappa(T)$ of
(Ba/Sr/Eu)$_8$Ga$_{16}$Ge$_{30}$ at low temperatures ($<$ 15 K) is
determined by scattering of phonons on charge carriers. Most
recently, Bridges and Downward proposed another possible mechanism
for the glasslike $\kappa(T)$ of clathrates.\cite{8} They argued
that off-center displacement of guest atoms is crucial for
understanding the glasslike behavior in $\kappa(T)$. Therefore, it
is important to investigate the origin of the glasslike $\kappa(T)$
in clathrate compounds using single crystals to exclude effects of
other factors, such as scattering at grain boundaries. The existence
of large number of clathrate compounds and the amenability of their
framework supply opportunities for us to bring the issue to a close,
and to find high performance thermoelectric materials among them.

Until today, most of the work on clathrate compounds has been
focusing on type-I clathrates of silicon,\cite{9,10}
germanium,\cite{5,11,12} and tin.\cite{13} In this paper, we present
a comprehensive study on single crystals of
Ba$_8$Ga$_{16}$Sn$_{30}$, which crystallizes in the
type-\setcounter{enumi}{8}\Roman{enumi} clathrate structure (SG:
\emph{cI}54).\cite{14,15} There are only two known members in this
family of clathrates. The other is the $\alpha$-phase of
Eu$_8$Ga$_{16}$Ge$_{30}$, which transforms to $\beta$-phase (type-I
clathrate) above 696 $^\circ$C.\cite{5,11} One of the structural
features for type-\setcounter{enumi}{8}\Roman{enumi} clathrate
compounds is that there is only one kind of polyhedral cage for the
guest atoms, differing from the two kinds of cage in both type-I and
type-II clathrates. In Ba$_8$Ga$_{16}$Sn$_{30}$, Ba atoms are
encapsulated in cages composed of 23 atoms of $E$ = (Ga, Sn), which
are derived from pentagonal dodecahedra composed of $E_{20}$ atoms.
The existence of small $E_8$ voids in the network of $E_{46}$ is
another feature of type-\setcounter{enumi}{8}\Roman{enumi}
clathrates. We were motivated to investigate the title compound in
detail by the few reports on it in literature with a scattering of
lattice parameters, different melting behaviors, and promising
thermoelectric properties. Recently, we succeeded in growing large
single crystals of Ba$_8$Ga$_{16}$Sn$_{30}$. The structural,
transport, and thermal properties are presented here.

\section{Experiment}
\subsection{Crystal growth and structure refinement}

Single crystals were grown from Sn-flux. High purity elements were
mixed in an atomic ratio of Ba:Ga:Sn = 8:16:60 in an argon
atmosphere glovebox. The mixture sealed in an evacuated and
carbonized silica tube was heated slowly to 1270 K and reacted for 5
hours. Then, it was cooled to room temperature in two steps: fast
cooled to 720 K at first and kept at this temperature for 12 hours,
then slowly cooled down to room temperature. The well-shaped
crystals of 10 mm in diameter with a shiny metallic luster were
separated from the molten Sn solvent by centrifuging. The crystals
are not sensitive to air and moisture. Polished surfaces of the
crystals were examined by the use of optical microscopy and Laue
x-ray reflection to confirm their homogeneous single-crystal nature.
The composition of the crystals was examined by electron probe
microanalysis (EPMA) with a JEOL JXA-8200 microanalyzer. The same
result with a composition of Ba$_8$Ga$_{16.0}$Sn$_{30.7}$ was
obtained on several crystals, which is nearly the ideal
stoichiometric composition. Because Carrillo-Cabrera \emph{et
al}.\cite{15} suggested a positive thermoelectric power of a Ga-rich
sample in a short report, we have tried to grow crystals in Ga-flux
with initial atomic ratio of Ba:Ga:Sn = 8:38:30. However, we
obtained the stoichiometric crystals again, so the stoichiometric
compound seems to be very stable. The structure was refined with a
Rigaku R-AXIS diffractometer and with synchrotron radiation powder
x-ray diffraction (XRD) from 100 to 390 K. The powder XRD experiment
was carried out by using a large Debye-Scherrer camera installed at
beam line BL02B2, SPring-8, Japan. The wavelength of the incident
x-ray was 0.735 \AA.

\subsection{Measurements of thermal and transport properties}
Differential thermal analysis (DTA) was performed from room
temperature to 1200 K with ruthenium as a standard. The electrical
resistivity (thermoelectric power) was measured with a homemade
cryostat from 3 (5) to 300 K, and measured from 100 to 500 K (550 K)
with a commercial MMR measurement system. The data obtained with the
two systems are in good agreement in the overlapped temperature
range. The Hall coefficient was measured under a magnetic field of 1
Tesla from 4 to 300 K.  Thermal conductivity was measured with a
steady-state method from 1.5 to 150 K with a homemade cryostat. The
measurements of specific heat were carried out from 2 to 300 K with
a PPMS (Quantum Design).

%***********************
\begin{figure}[t]
\includegraphics[width=0.9\linewidth]{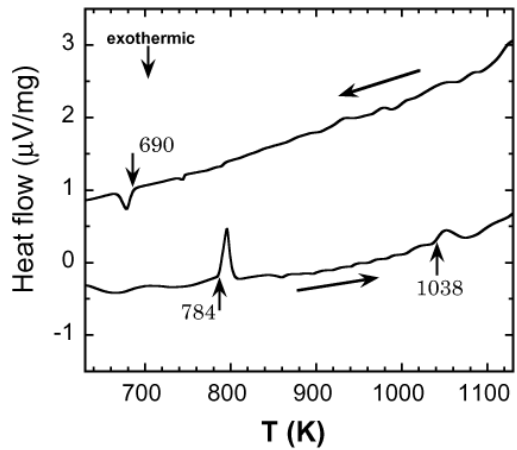}% Here is how to import EPS art
\caption{Differential thermal analysis curves for
Ba$_8$Ga$_{16}$Sn$_{30}$.} \label{fig1}
\end{figure}
%***********************

\section{Results and discussion}
It was reported at first that Ba$_8$Ga$_{16}$Sn$_{30}$ melts
congruently at 723 K,\cite{14} whereas, Kuznetsov \emph{et
al}.\cite{16} observed an incongruent melting behavior with a
decomposition temperature of 740 K and liquidus temperature of 784
K. In order to check which is the case, a DTA measurement was
carried out on our small single crystals. Figure 1 shows the heating
and cooling DTA curves. The double peaks on the heating curve
indicate an incongruent melting nature of this compound. However, we
observed a decomposition temperature of 784 K and liquidus
temperature of 1038 K, which are higher than those reported in
Ref.16. This information helped us to succeed in growing large
single crystals of Ba$_8$Ga$_{16}$Sn$_{30}$ by cooling down the
reactant in two steps as described above.

%*********************************************
\begin{figure}[t]
\includegraphics[width=0.85\linewidth]{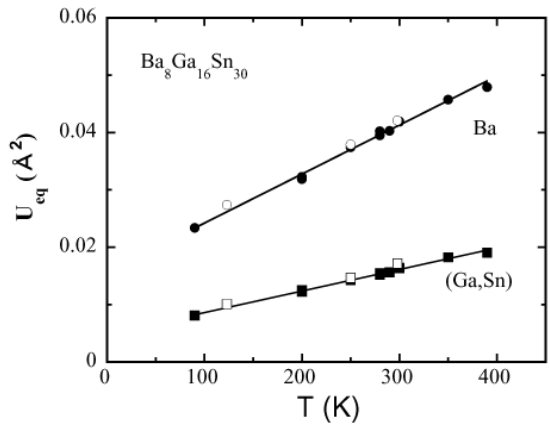}% Here is how to import EPS art
\caption{Temperature dependence of isotropic atomic displacement
parameters $U_{eq}$ of Ba$_8$Ga$_{16}$Sn$_{30}$. Open symbols: data
determined by structural refinement from single crystal XRD. Closed
symbols: data determined by structural refinement from synchrotron
radiation powder XRD.} \label{fig2}
\end{figure}

The single-crystal XRD data were collected at 293, 253 and 123 K,
respectively. The cubic crystal structure of
type-\setcounter{enumi}{8}\Roman{enumi} clathrate for
Ba$_8$Ga$_{16}$Sn$_{30}$ was confirmed. The lattice parameter
decreases from 11.586(1) to 11.5831(4) to 11.5619(3) \AA \ with
decreasing temperature correspondingly. The structural refinement
gives a composition of Ba$_8$Ga$_{16.2}$Sn$_{29.8}$, which is in
good agreement with the results of EPMA. The high resolution powder
XRD analysis using synchrotron radiation gives consistent results
also. With decreasing temperature from 390 to 100 K, there is a
normal thermal contraction, but no change in structure was observed.
There is no indication of split sites for Ba atoms down to 100 K.

Here, we pay attention to the atomic displacement parameters (ADPs)
obtained from XRD analysis. As a first approximation, the guest
atoms in clathrate compounds may be treated as Einstein oscillators,
which vibrate independently with the same frequency, and the
framework atoms as a Debye solid. It has been proved to be
successful in giving a reasonable estimation of the Einstein
temperature $\Theta_{\rm E}$, Debye temperature $\Theta_{\rm D}$ and
room-temperature thermal conductivity with ADPs of the guest and
framework atoms.\cite{17} Figure 2 shows the temperature dependences
of the isotropic ADPs for Ba$_8$Ga$_{16}$Sn$_{30}$. The open and
closed symbols denote the results from single-crystal XRD analysis
and synchrotron radiation powder XRD analysis, respectively, which
are in good agreement. The ADPs of Ba atoms are much larger than
those of framework atoms (Ga,Sn), which indicates the rattling of Ba
atoms in the oversized cages as in the
Ba$_8$Ga$_{16}$Ge$_{30}$.\cite{5} The two straight lines, which do
not pass through the origin by extrapolation, are linear fits of the
data. Using the slopes of these straight lines, $U_{eq}/T$ =
$h^2/(4\pi^2m_{r}k_{\rm B}\Theta_{\rm E}^2)$ and $U_{eq}/T$ =
$3h^2/(4\pi^2m_{av}k_{\rm B}\Theta_{\rm D}^2)$ (where $m_r$ and
$m_{av}$ are mass of the rattler atom and average mass of the
framework atoms, respectively), $\Theta_{\rm E}$ = 64 K and
$\Theta_{\rm D}$ = 195 K for Ba$_8$Ga$_{16}$Sn$_{30}$ were obtained.
As discussed below, they are close to the two characteristic
temperatures extracted from specific heat data. Because there are no
evident split sites in Ba$_8$Ga$_{16}$Sn$_{30}$ from XRD analysis,
the Ba atoms in this compound are normal rattlers like in
Ba$_8$Ga$_{16}$Ge$_{30}$. The large ADP values describe rattling of
Ba atoms around the centers of their crystallographic sites.
%*********************************************
\begin{figure}[t]
\includegraphics[width=0.85\linewidth]{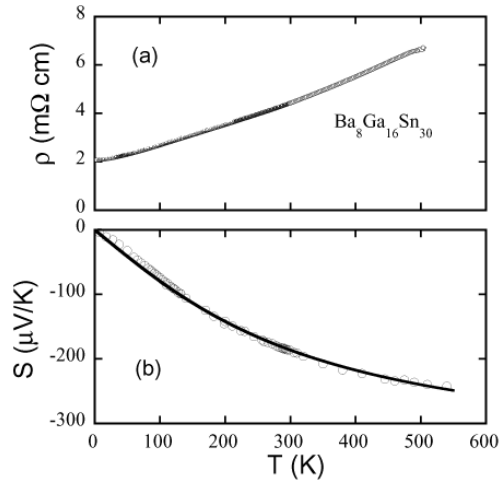}% Here is how to import EPS art
\caption{(a) Temperature dependence of electrical resistivity $\rho$
and (b)thermoelectric power $S$ of Ba$_8$Ga$_{16}$Sn$_{30}$. The
solid line is a calculation result (see text).} \label{fig3}
\end{figure}
%***********************************

The temperature dependence of the electrical resistivity $\rho$ and
thermoelectric $S$ of Ba$_8$Ga$_{16}$Sn$_{30}$ is shown in Fig. 3.
On cooling, $\rho(T)$ decreases monotonically from 6.6 m$\Omega$ cm
(500 K) to 2.0 m$\Omega$ cm (3 K), being typical of heavily doped
semiconductors. Our observation is contrasting with the results in
Ref.16, where a typical semiconductor  behavior with a value of 15
meV for activation energy was observed. The carrier concentration
$n$ (2.2$\times$10$^{19}$ cm$^{-3}$) of their polycrystalline sample
at room temperature is a little smaller than that of our
single-crystal sample $n$(300 K) = 3.7$\times10^{19}$ cm$^{-3}$ (see
below). This difference might be responsible for the different
behavior in $\rho(T)$. Metal-like temperature dependence of $\rho$
was reported for a Ga-rich polycrystalline sample.\cite{15}

The absolute value of $S$ increases monotonically with increasing
temperature up to 550 K. The overall features of the $S(T)$ resemble
the previously reported results in Ref.16. However, the maximum at
about 500 K reported in Ref.16 does not exist in our data in Fig.
3(b). The discrepancy might result from the possibility that our
accessible temperature was not high enough to observe a maximum or
from distinct quality between our single-crystal sample and their
polycrystalline sample.

As the detailed energy band structure of Ba$_8$Ga$_{16}$Sn$_{30}$ is
not known yet, the $S(T)$ and band effective mass $m^*$ are
evaluated by an assumption of one parabolic conduction band model
with different scattering mechanisms. The overall feature of $S(T)$
could be reproduced well by a single-band model with dominant
ionized impurity scattering. In this model the thermoelectric power
and the carrier concentration is given by\cite{18}

\begin{equation}
S(T)= \frac{k_B}{e}(\frac{4F_3(\eta)}{3F_2(\eta)}-\eta)
\end{equation}
\begin{equation}
n = \frac{(2m^*k_BT)^{3/2}}{2\pi^2\hbar^3}F_{1/2}(\eta)
\end{equation}
where $F_x$ is Fermi-Dirac integral of the order $x$, $\eta$ is
reduced Fermi energy defined as $\eta$= $E_{\rm F}/k_{\rm B}T$
($E_{\rm F}$ is Fermi energy).

\begin{figure}[t]
\includegraphics[width=0.85\linewidth]{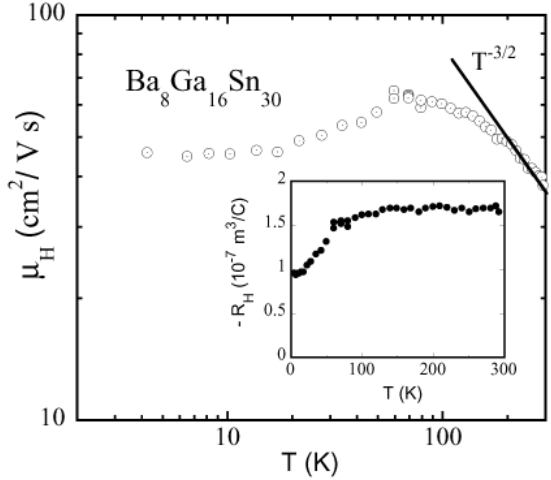}% Here is how to import EPS art
\caption{Temperature dependence of Hall mobility $\mu_{\rm H}$ of
Ba$_8$Ga$_{16}$Sn$_{30}$. Inset shows temperature dependence of
Hall coefficient $R_{\rm H}$ of Ba$_8$Ga$_{16}$Sn$_{30}$.}
\label{fig4}
\end{figure}

The solid line in Fig. 3(b) is the calculated $S(T)$ with $E_{\rm
F}$ = 88 meV, which reproduces our experimental data very well. The
estimated band effective mass $m^*$ = 0.14$m_0$ ($m_0$ is the free
electron mass) was obtained by using the Fermi energy and the room
temperature carrier concentration. The $m^*$ value is one order of
magnitude smaller than 3.6$m_0$ for the type-I clathrate
Ba$_8$Ga$_{16}$Ge$_{30}$ estimated by the similar method.\cite{7}
Recently, band structure calculation was reported for Eu filled
type-\setcounter{enumi}{8}\Roman{enumi} clathrate of germanium,
which suggests that the low band effective mass might be inherent to
$n$-doped type-\setcounter{enumi}{8}\Roman{enumi} clathrates due to
their structural features.\cite{19} The disperse bands centered
around the $E_8$ voids in \setcounter{enumi}{8}\Roman{enumi}
clathrates would be responsible for the low $m^*$. However, there is
not enough experimental data to examine whether it is inherent or
not because the member of type-\setcounter{enumi}{8}\Roman{enumi}
clathrates is limited to 2 at present.

In order to get further insights into the carrier scattering
mechanisms in Ba$_8$Ga$_{16}$Sn$_{30}$, the Hall coefficient $R_{\rm
H}$ was measured from 4 to 300 K. As shown in Fig. 4, $R_{\rm H}$ is
negative in the overall temperature range. The negative $S$ and
$R_{\rm H}$ over a wide temperature range indicate the majority
carriers being electrons in Ba$_8$Ga$_{16}$Sn$_{30}$. Assuming a one
band model, the carrier concentration $n$ (= 1/$eR_{\rm H}$) is
derived to be 3.7$\times$10$^{19}$cm$^{-3}$ at 300 K, which is
increased to 6.6$\times$10$^{19}$cm$^{-3}$ at 4 K. The Hall mobility
$\mu_{\rm H}$ = $|R_{\rm H}|$/$\rho$ is plotted in the inset of
Fig.4 as a function of temperature. At room temperature, $\mu_{\rm
H}$ = 39 cm$^2$/V s, is larger than 20 cm$^2$/V s of
$\alpha$-Eu$_8$Ga$_{16}$Ge$_{30}$ (Ref.11) and 26 cm$^2$/V s of a
Ba$_8$Ga$_{16}$Sn$_{30}$ polycrystalline sample.\cite{16} In the
relaxation time approximation, temperature dependence of $\mu_{\rm
H} \propto T^\alpha$ determines the carrier scattering mechanism:
$\mu_{\rm H}$ taking the values of 3/2, 0, -3/2 for ionized
impurity, neutral impurity, and acoustic phonon scattering,
respectively.\cite{20} However, it is difficult to observe the ideal
power law of $\mu_{\rm H}$ experimentally over a wide temperature
range in a real solid. Instead, a mixed scattering process of ion
impurity and acoustic phonon scattering is usually
observed.\cite{21} For the present compound, $\mu_{\rm H}$ shows
weak temperature dependence below 20 K, increases with increasing
temperature between 20 and 70 K, then decreases above 80 K. An
approximately $T^{-3/2}$ dependence was observed near 300 K. This
temperature dependence of $\mu_{\rm H}$ indicates a crossover from
dominant charge carrier scattering by neutral impurities below 20 K
to acoustic phonon scattering at higher temperature via an ionized
impurity scattering range.

\begin{figure}[t]
\includegraphics[width=0.85\linewidth]{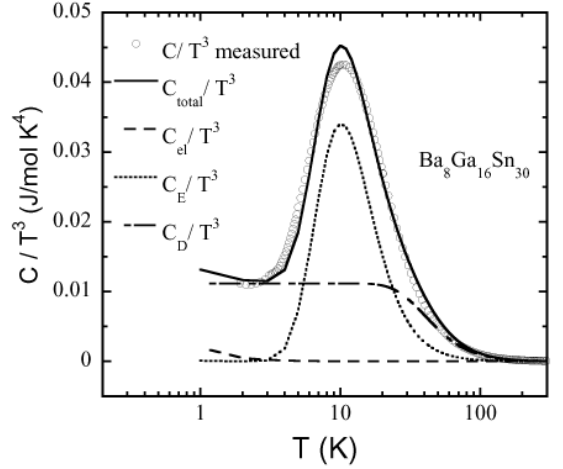}% Here is how to import EPS art
\caption{Temperature dependence of specific heat $C$ of
Ba$_8$Ga$_{16}$Sn$_{30}$. The lines are results of fitting(see
text).} \label{fig5}
\end{figure}
  As mentioned earlier, in first approximation, the Ba atoms could
  be considered as Einstein oscillators and the framework composed of
  (Ga,Sn)$_{46}$ clusters as a Debye solid. Following this approach, the specific
  heat of Ba$_8$Ga$_{16}$Sn$_{30}$ is treated as a sum of three terms: an electronic
  contribution $C_{el}$, a Debye contribution $C_{\rm D}$, and an Einstein contribution $C_{\rm
  E}$
  with $\Theta_{\rm E}$ of the order of several tens Kelvin. The low-energy vibrating
   modes would greatly contribute to low-temperature specific heat.
   To elucidate the evidence for the low-energy modes in Ba$_8$Ga$_{16}$Sn$_{30}$,
   the specific heat was measured from 2 to 300 K. In order to emphasize
    the contribution of the local modes, the data are shown in Fig. 5
    as a plot of $C/T^3$ vs $T$. It can be seen clearly that a broad peak
    centered at 10 K exists. In the $C/T^3$ vs $T$ plot, the Debye contribution
    approaches a constant at low temperatures. For semiconducting
    clathrate compounds, the electronic contribution to specific heat is
    a small portion of the total specific heat at low temperatures and becomes
    relatively smaller with increasing temperature. Therefore, the peak mainly
    comes from the local vibrating modes. With this analysis, we fit the data to
    an expression of specific heat $C/T^3$ = $\gamma/T^2$ + $N_{\rm E}C_{\rm E}/T^3$
    + $N_{\rm D}C_{\rm D}/T^3$. With the electronic specific heat
     coefficient $\gamma$ of 1.3 mJ/molK$^2$ obtained from the low-temperature plot of
     $C(T)/T$ vs $T^2$, we further fixed the numbers of Debye and Einstein
     oscillators to $N_{\rm E}$ = 8 and $N_{\rm D}$ = 46, respectively, which are the numbers of
     guest Ba atoms and framework atoms of (Ga,Sn) per formula unit. Then the
     fitting parameters are just the two characteristic temperatures $\Theta_{\rm D}$ and
    $\Theta_{\rm E}$. The fitting results for the three contributions and their sum
     $C_{total}$ are shown together in Fig. 5. The two parameters obtained from the
     fitting are $\Theta_{\rm D}$ = 200 K and $\Theta_{\rm E}$ = 50 K. They are close to the
     values of 195 K and 64 K estimated with ADPs. Considering the simplicity of the model,
     the fit with two parameters is fairly good. Better agreement could be
     achieved by assuming a distribution of $\Theta_{\rm E}$ like the approach in the
     analysis of specific heat data of ZrW$_2$O$_8$.\cite{22} Furthermore, coupling
     effects between the local modes of guest atoms and low-frequency acoustic
     phonons of the framework atoms should be taken into account.

\begin{figure}[t]
\includegraphics[width=0.85\linewidth]{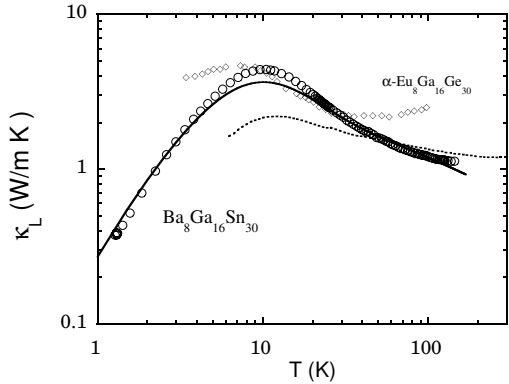}% Here is how to import EPS art
\caption{Temperature dependence of lattice thermal conductivity
$\kappa_{\rm L}$ of Ba$_8$Ga$_{16}$Sn$_{30}$. The solid line is a
fit of data (see text). The data of $\kappa_{\rm L}$ for $\alpha$-
Eu$_8$Ga$_{16}$Ge$_{30}$($\diamond$) and a polycrystalline sample
of Ba$_8$Ga$_{16}$Sn$_{30}$ (dashed line) were taken from Ref. 11
and Ref. 23, respectively.} \label{fig6}
\end{figure}

It is believed that the scattering of heat-carrying acoustic phonons
of the framework atoms by the local modes of the guest atoms is
responsible for the reduction of thermal conductivity. The lattice
thermal conductivity $\kappa_{\rm L}$ of Ba$_8$Ga$_{16}$Sn$_{30}$ is
plotted as a function of temperature in Fig. 6. For comparison,
previously reported data of $\kappa_{\rm L}(T)$ are also shown for
polycrystals of Ba$_8$Ga$_{16}$Sn$_{30}$ (Ref. 23) and
$\alpha$-Eu$_8$Ga$_{16}$Ge$_{30}$ (Ref. 11) At low temperatures, the
$\kappa_{\rm L}(T)$ of our single crystal is larger than that of the
polycrystalline sample,\cite{23} in which scattering of phonons at
grain boundaries might greatly contribute to the reduction of
$\kappa_{\rm L}(T)$. The $\kappa_{\rm L}(T)$ of the two
type-\setcounter{enumi}{8}\Roman{enumi} clathrate compounds is
characterized by a peak at about 10 K. A significant peak in
$\kappa_{\rm L}(T)$ is typical of a crystalline solid, differing
from the glasslike $\kappa_{\rm L}(T)$ observed for
(Sr/Eu)$_8$Ga$_{16}$Ge$_{30}$ type-I clathrates.\cite{5} Below the
temperature of the peak, $\kappa_{\rm L}$ of
Ba$_8$Ga$_{16}$Sn$_{30}$ decreases faster than that of
$\alpha$-Eu$_8$Ga$_{16}$Ge$_{30}$ with decreasing temperature.

A qualitative understanding of the contributions from different
scattering mechanisms to $\kappa_{\rm L}$ could be reached by
analysis of $\kappa_{\rm L}(T)$ data with a phenomenological
model.\cite{24,25,11} In this model, the lattice thermal
conductivity is formulated as following:
\begin{equation}
\kappa_{\rm L} = \frac {\nu}{3}\int_0^{\omega_{\rm
D}}C(\omega)l(\omega)d\omega
\end{equation}
with $l=(l^{-1}_{TS} + l^{-1}_{Res} + l^{-1}_{R})^{-1} + l_{min}$,
     $l^{-1}_{TS}=A(\hbar\omega/k_{\rm B})\tanh(\hbar\omega/2k_{\rm
     B}T)+(A/2)(k_{\rm B}/\hbar\omega+B^{-1}T^{-3})^{-1}$,
     $l^{-1}_{Res}=\sum
     C_i\omega^2T^2/[(\omega_i^2-\omega^2)^2-\gamma_i(\omega_i\omega)^2]$,
     and $l^{-1}_{R}=D(\hbar\omega/k_{\rm B})^4$, where $\nu$ is the average
velocity of sound, $C(\omega)$ is the Debye specific heat, and
$l(\omega)$ is the total mean free path of phonons with frequency of
$\omega$. The three components of $l(\omega)$ correspond to the
contributions from different scattering mechanisms: tunneling states
$(l_{TS})$, Reyleigh scattering $(l_R)$, and resonant scattering
$(l_{Res})$. The lower limit of $l(\omega)$ is constrained to a
constant $l_{min}$. We followed the approach in Ref. 11 to reduce
the number of fitting parameters. Two parameters, i.e., the velocity
of sound $\nu = (\Theta_{\rm D}k_{\rm B}/\hbar)/(6\pi^2n_A)^{1/3}$
($n_A$ is the number of atoms per unit volume) and resonant
frequency $\omega_{\rm E} = k_{\rm B}\Theta_{\rm E}/\hbar$ were
fixed, respectively, by use of the Debye temperature $\Theta_{\rm
D}$ = 200 K and Einstein temperature $\Theta_{\rm E}$ = 50 K
obtained from the experiments of specific heat. Other parameters
($A, B, C, D,\gamma_1$) were obtained by fitting the data to the
model. The solid line in Fig. 6 shows a fit with a set of reasonable
parameters: $A$ = 1.08$\times$10$^4$ m$^{-1}$K$^{-1}$, $B$ =
5.0$\times$10$^{-1}$K$^{-2}$, $C$ = 1.0$\times$10$^{30}$ m$^{-1}$
s$^{-2}$ K$^{-2}$, $D$ = 2.6 m$^{-1}$ K$^{-4}$, $\gamma_1$ = 0.8.
The ratio of $A/B$ is a measure of the density of tunneling states
per unit volume strongly coupled to phonons.\cite{24} Here, we
obtained $A/B$ = 2.2$\times$10$^4$ m$^{-1}$K$^{-3}$. This ratio is
comparable to 3.7$\times$10$^4$ m$^{-1}$K$^{-3}$ of
$\alpha$-Eu$_8$Ga$_{16}$Ge$_{30}$
(type-\setcounter{enumi}{8}\Roman{enumi}), but much smaller than
3.6$\times$10$^6$ m$^{-1}$K$^{-3}$ for
$\beta$-Eu$_8$Ga$_{16}$Ge$_{30}$ (type-I), which shows glasslike
$\kappa_{\rm L}(T)$.\cite{11} It is suggestive that the density of
tunneling states is very low in the
type-\setcounter{enumi}{8}\Roman{enumi} clathrates if any. It is
straightforward to understand if one attribute the glasslike
$\kappa_{\rm}$ $L(T)$ to the tunneling of guest atoms among the
split sites, which are absent in
type-\setcounter{enumi}{8}\Roman{enumi} clathrates. Massive and
smaller guest atom, such as Eu and Sr, filled
type-\setcounter{enumi}{8}\Roman{enumi} clathrate of tin is expected
to show glasslike thermal conductivity because Eu/Sr atoms would
have split sites or have much more room to move around in cages if
off-center displacement of guest atoms is really responsible for
glasslike thermal conductivity as suggested in Ref. 8. The stability
of Sr filled type-\setcounter{enumi}{8}\Roman{enumi} clathrate
Sr$_8$Ga$_{16}$Ge$_{30}$ is predicted from band
calculation,\cite{19} but has not been confirmed experimentally.

A dimensionless figure of merit $ZT$ = 0.15 at 300 K for
Ba$_8$Ga$_{16}$Sn$_{30}$ is estimated from the present set of data.
For thermoelectric application, the $ZT$ should be improved by
optimal doping level and further reduction of thermal conductivity.
As mentioned above, a massive and smaller guest atom, such as Eu,
filled type-\setcounter{enumi}{8}\Roman{enumi} clathrate is expected
to have lower thermal conductivity. Furthermore, band structure
calculation on type-\setcounter{enumi}{8}\Roman{enumi} Ge-clathrate
suggests that $p$-doped type-\setcounter{enumi}{8}\Roman{enumi}
clathrate is promising for thermoelectric application, for which a
figure of merit of 1.2 at 400 K is predicted. Therefore, it is
interesting to fabricate and study $p$-doped
type-\setcounter{enumi}{8}\Roman{enumi} clathrate compounds.

\section{Summary}
Single crystals of type-\setcounter{enumi}{8}\Roman{enumi} clathrate
compound Ba$_8$Ga$_{16}$Sn$_{30}$ were grown from Sn-flux.
Incongruent melting nature of this compound was confirmed by
differential thermal analysis. Negative thermoelectric power and
Hall coefficient indicate electrons dominating the transport
properties. The estimated band effective mass 0.14$m_0$ is smaller
than that of type-I clathrate compounds. The large absolute value of
thermoelectric power (188 $\mu$V/K at 300 K) may originate from the
low carrier concentration $n$(300 K) = 3.7$\times$10$^{19}$
cm$^{-3}$. Hall mobility $\mu_{\rm H}$ shows a maximum of 62
cm$^2$/V s around 70 K. The analysis of the temperature dependence
of $\mu_{\rm H}$ suggests a crossover of dominant scattering
mechanism from ion impurity at low temperatures to acoustic phonon
scattering at high temperatures. Although the $\kappa(T)$ shows a
pronounced peak, being typical of crystalline solids, the value of
thermal conductivity is reduced very much. $\kappa$ = 1.1 W/m K at
150 K. The reduction in $\kappa(T)$ is attributed to the rattling of
Ba atoms in the cages composed of Ga and Sn atoms. The evidence of
this rattling is elucidated by the analysis of experimental data of
XRD and specific heat, which gives the estimation of $\Theta_{\rm
D}$ = 200 and $\Theta_{\rm E}$ = 50 K, respectively. It is
interesting to study $p$-doped
type-\setcounter{enumi}{8}\Roman{enumi} clathrate compounds to
examine the predictions of band structure calculations that these
compounds should have prospective thermoelectric properties.

 \vskip 10mm
%\begin{acknowledgments}
D.H. acknowledges the financial support from JSPS. We thank Y.
Shibata for the electron-probe microanalysis. This work was
supported by a Grant-in Aid for Scientific Research (COE 13CE2002)
and a Grant-in-Aid for Scientific Research in Priority Area
"Skutterudite"(No.15072205) of MEXT Japan. The synchrotron x-ray
diffraction was performed at the BL02B2 in SPring-8 under Proposal
No. 2003A0247.
%\end{acknowledgments}

%\newpage

\end{document}